\documentclass[twocolumn]{aastex7}
\usepackage{xcolor}
\usepackage{amsmath}
\usepackage{gensymb}
\usepackage{makecell}
\usepackage{soul}

\begin{document}

\newcommand{\vdag}{(v)^\dagger}
\newcommand\aastex{AAS\TeX}
\newcommand\latex{La\TeX}
\newcommand{\tmo}[1]{{\textcolor{teal}{talia: #1}}}

\newcommand{\addedit}[1]{{#1}}
\renewcommand\st[1]{}
\newcommand{\removeedit}[1]{\textcolor{red}{\st{#1}}}

\newcommand{\Msun}{\mathrm{M_\odot}}

\title{Shooting for the stars: Jet-mode feedback and AGN jet deceleration from stellar mass-loading}

\author[orcid=0009-0006-0548-9855]{Talia M. O'Shea}
\affiliation{Department of Astronomy,  University of Wisconsin--Madison, 475 N.~Charter St., Madison, WI 53706, USA}
\email[show]{tmoshea@wisc.edu}
\author[orcid=0000-0002-8433-8652]{Sebastian Heinz}
\affiliation{Department of Astronomy,  University of Wisconsin--Madison, 475 N.~Charter St., Madison, WI 53706, USA}
\email{sheinz@wisc.edu}
\author[orcid=0000-0001-7493-7419]{Melinda Soares-Furtado}
\affiliation{Department of Astronomy,  University of Wisconsin--Madison, 475 N.~Charter St., Madison, WI 53706, USA}
\affiliation{Department of Physics, 2320 Chamberlin Hall, University of Wisconsin--Madison, 1150 University Avenue Madison, WI 53706-1390}
\email{mmsoares@wisc.edu}
\author[orcid=0000-0001-9274-1145]{Zsofi Igo}
\affiliation{Max-Planck-Institut für Extraterrestrische Physik (MPE), Giessenbachstrasse 1, 85748 Garching bei München, Germany}
\affiliation{Exzellenzcluster ORIGINS, Boltzmannstr. 2, 85748, Garching, Germany}
\email{zigo@mpe.mpg.de}
\author[orcid=0000-0002-0761-0130]{Andrea Merloni}
\affiliation{Max-Planck-Institut für Extraterrestrische Physik (MPE), Giessenbachstrasse 1, 85748 Garching bei München, Germany}
\email{am@mpe.mpg.de}

\begin{abstract}
AGN jets are thought to be vital ingredients in galaxy evolution through the action of kinetic feedback; however, how narrow, relativistic outflows couple to galaxies remains an open question. Jet deceleration, which is often attributed to the entrainment of material, such as stellar winds, is thought to be necessary for efficient coupling. We present a simple model of jet deceleration due to stellar mass-loading to investigate the energy budget of direct jet feedback in the local Universe. To this end, we produce models of stellar mass-loss, including deriving a prescription for main sequence mass-loss rates as a function of stellar population age. We pair this mass-loss data with a parametric fit for radio AGN incidence, predicting that a majority of jets are decelerated within their hosts, and generally replicate the expected FR-II fraction in LERGs. We calculate that $\gtrsim$25\% of the jet power in the local Universe is efficiently decelerated and available for direct feedback within galaxies for any stellar population age. This fraction is largely invariant to the shape of the radio AGN incidence function at low jet Eddington fractions. The stellar mass-loss rate evolves significantly over time, approximately following $\tau^{-1.1}$, leading to corresponding decreases in decelerated jet power in older stellar populations. Although asymptotic giant branch (AGB) stars dominate mass-loss at all ages, we find that their stochasticity is important in low-mass galaxies, and derive a critical jet power below which main sequence stars alone are sufficient to decelerate the jet.
\end{abstract}


\section{Introduction}\label{sec:intro}

\setcounter{footnote}{0}%
In recent decades it has become apparent that supermassive black holes (SMBH) profoundly impact the evolution of their host galaxies, and vice versa \citep[\removeedit{e.g., the M-$\sigma$ relation;}\addedit{see e.g.,}][\addedit{and references within}]{Hardcastle2020, Hlavacek2022, Mukherjee2025}. Galaxy properties and environment (e.g., cool gas reservoirs) shape the accretion flows onto SMBHs, producing active galactic nuclei (AGN). AGN then couple with the rest of the galaxy \addedit{and the surrounding environment} through a variety of mechanisms collectively referred to as feedback. 

\addedit{Feedback from AGN comes in many forms \citep[see e.g.,][]{Ishibashi2014, Wang2024}. In the local Universe, radio-loud AGN are typically hosted by AGN} with geometrically thick but optically thin accretion disks, which are thought to accrete inefficiently through advection-dominated accretion flows \citep[][]{Narayan1994, Narayan1995}. \addedit{Radio-loud AGN are generally associated with relativistic AGN jets.} Though the exact mechanisms of jet launching remain under study \citep[see e.g.,][]{Janiuk2022}, magnetic fields surrounding the SMBH are thought to launch collimated jets of charged particles \addedit{\citep{Blandford1977, Begelman1984}}, \addedit{with the synchrotron emission from these jets observed in radio wavelengths.} \removeedit{The synchrotron emission from these jets can be seen in radio wavelengths, so they are often called radio-loud AGN.  }

\removeedit{Feedback from AGN can be broadly characterized by the accretion state of the AGN.} \removeedit{quasar mode, the AGN} \addedit{Yet other AGN} accrete efficiently in geometrically thin, but optically thick, Shakura-Sunyaev disks \citep[][]{Shakura1973}. The turnover between these two accretion states is thought to occur at Eddington ratios of a few percent \citep[e.g.,][]{Merloni2008}. \removeedit{AGN are not just differentiated by their accretion mode, but also at larger scales, where low accretion rate AGN frequently produce relativistic jets.}Radio-loud AGN hosted in radiative mode SMBH, also known as high excitation radio galaxies (HERGs) and considered to be in so-called radiative mode, typically produce excitation lines that can be identified and characterized by their optical spectra \citep[see e.g.,][]{Kauffmann2003, Bongiorno2010}. In the local Universe, \addedit{few jets are hosted by efficiently accreting HERGs}\removeedit{jet-mode AGN accrete efficiently or exhibit strong emission lines (i.e., HERGs)}, instead being predominately hosted in low excitation radio galaxies \citep[LERGs;][]{Kauffmann2008, Weigel2017, Kondapally2022, Mingo2022}.
 
Locally, the host galaxy populations of AGN also vary widely. For example, X-ray surveys have found enhanced AGN fractions among blue and green galaxies \citep{Aird2012}. Radio AGN, however, are more frequently found on the red sequence \citep[see e.g.,][]{Best2005,Hickox2009, Janssen2012, Ching2017, Williams2018}, in massive elliptical galaxies \citep[e.g.,][]{Delvecchio2022}, though an increasing number of radio AGN are found in bluer, star-forming galaxies \citep{Tadhunter2005, Kondapally2022, Igo2024}. 

Further complicating the picture, radio loud AGN are not a homogeneous class of sources. Aside from the division between LERGs and their counterpart HERGs, resolved jets can be separated morphologically into Fanaroff-Riley \citep[FR;][]{Fanaroff1974} classes, with edge- or lobe-brightened sources composing FR-II jets and centrally brightened sources known as FR-Is. \addedit{FR-I jets are decelerated and decollimated on kpc scales \citep[see e.g.,][]{Laing2002, Laing2014}, while FR-II jets remain relativistic until they produce termination shocks, also referred to as working surfaces or hot spots.} FR-IIs generally have higher radio luminosities \citep[$L_{\rm 150~MHz} \gtrsim10^{33} ~\rm erg~s^{-1}~Hz^{-1}$, see e.g.,][]{Mingo2019} than FR-I sources, though there is a region of overlap in radio power (see Sect. \ref{sec:disc_morph}, and references therein). Recent work has also highlighted a growing number of so-called FR-0 compact sources, which are lower-power than FR-I jets \citep[see e.g.,][]{Baldi2015, Baldi2023}.

Much of the power from jet-mode AGN is deposited into these kinetic outflows, as opposed to radiative power \citep{Churazov2005}. Such AGN have long been known to interact with the \addedit{circumgalactic medium (CGM) and intergalactic medium (IGM) surrounding their hosts}, i.e. through X-ray observations of shocks and bubbles produced by radio lobes \citep{Boehringer1993, Fabian2003, Hardcastle2020, Hlavacek2022}. In galaxy clusters, jets are thought to prevent cooling of the intracluster medium \citep[ICM;][]{mcnamara_heating_2007}. 

At redshifts $z \leq 2$, \citet{Kondapally2023} found that the kinetic energy density of radio-loud AGN galaxies is dominated by AGN jets, rather than winds or star formation, and \cite{Buchner2024} further showed that AGN jets are the largest impediment to galaxy growth for massive galaxies in the local Universe. There is significant power in these jets, with work by {\cite{Best2006, Smolcic2017, Hardcastle2019}} finding that mechanical heating from AGN closely balanced cooling rates in elliptical galaxies \addedit{at low redshifts}, and indicating that mechanical feedback is essential to the star-forming evolution of such hosts. Thus, AGN jets are crucial to \addedit{models of AGN feedback.}\removeedit{understanding how AGN couple with the interstellar medium (ISM) of their host galaxies} 

\removeedit{In particular, we are interested in how jet kinetic energy is converted into thermal energy, preventing the resumption of star formation in red sequence host galaxies. This process remains an active area of study. Indeed, not all jets are thought to deposit a significant amount of energy into the ISM of their host galaxies.} 

\addedit{Jets inject energy into haloes on a wide range of scales. Early work primarily focused on the impact of AGN jets on the outer halo, with jet kinetic energy heating the CGM or ICM, maintaining a hot halo, and preventing cooling flows (often called preventative feedback). Recently, there has been growing interest in the impact of jets on the inner kpcs of their hosts \citep[what we call ``direct'' feedback, see e.g.,][and references therein]{Mukherjee2025}.} However, numerical work has found that AGN jets can clear the interstellar medium (ISM), creating a low-density passage that the jet traverses easily, such that energy is not deposited into the host galaxy \citep[e.g.,][]{Vernaleo2006}. To couple the jet and host galaxy, the jet must have an increased effective opening area, and/or it must be decelerated so that the initially relativistic jet material interacts with the ISM. \addedit{There are several potential ways to produce a large jet opening area, such as a large precession angle \citep[][]{Heinz2006, Nawaz2016, Cielo2018, Su2021}, light jets \citep{Guo2016, Su2021, Ehlert2023, Weinberger2023}, cluster weather \citep{Heinz2006, Bourne2021, YatesJones2023}, and interactions with the ISM \citep{Mukherjee2018}. Once decelerated, some fraction of the jet's kinetic energy will heat the surrounding medium \citep[see e.g.,][]{Wagner2012, Mukherjee2016}, a later step in the feedback process that we do not focus on here.}

Jet deceleration is believed to come from the incorporation, \addedit{called mass-loading or entrainment,} of material into the jet. Momentum conservation forces the jet to slow as the mass becomes entrained and accelerated to jet speed, thus allowing the jet to broaden and couple with the \addedit{surrounding gas}\removeedit{ISM}. Deceleration can occur via turbulent mixing at the jet boundary \citep{bicknell_model_1984} or the entrainment of stellar wind material \citep{Phinney1983,Komissarov1994}, as modeled in the detailed models of jet deceleration in \cite{Laing2002}. \addedit{In this work, we focus on the contributions of stellar winds to mass-loading. Therefore, we consider a ``galaxy" to be primarily the stellar component of a halo, extending to ${\sim}$10s of kpc. We seek to understand under what conditions jets are decelerated by the stellar winds of their hosts, and thus differentiate between decelerated jets that may couple to the ISM from undecelerated jets that couple with the CGM or ICM.}

\addedit{The relative contributions of stellar winds and the ISM to mass-loading are not precisely known. There are many studies of jet propagation and energy injection into the ISM \citep[see e.g.,][]{Wagner2012, Mukherjee2016, YatesJones2021, Bhattacharjee2024, Borodina2025} with varying inclusion of density profiles, turbulence, magnetic fields, and more. Some numerical works have suggested that the ISM is the dominant mass-loading source for powerful jets \citep[e.g.,][]{Perucho2014}. However, while the ISM can be driven out of the jet path by the cocoon over the jet lifetime \citep[][]{Vernaleo2006, Wagner2012}, the presence of stellar winds should remain generally unaffected.}

The analytical model developed by \cite{hubbard_active_2006} has suggested that the presence of even one star of exceptionally high mass-loss rate (MLR; e.g., a Wolf-Rayet star) could effectively decelerate an AGN jet.  Hydrodynamic simulations of mass-loading have found that an old stellar population could decelerate jets of kinetic power ${\sim} 10^{42-43}~\rm erg~s^{-1}$ \citep[][]{Perucho2014}. Although jets can have powers up to ${\sim}10^{48}~\rm erg~s^{-1}$, a significant amount of jet energy in the local Universe is hosted in less powerful, but more common, radio AGN \citep[][]{Merloni2008, Weigel2017, Baldi2023}.

Until now, most simulations of mass-loading in jets begin with a mass-loading rate (e.g., in $\rm g~cm^{-3}~s^{-1}$). Here we seek to complement such numerical work, introducing in Section \ref{sec:meth_jet} a revised analytical approach similar to \cite{hubbard_active_2006}, accounting for the time evolution of mass-loss as stellar populations age.\removeedit{In Sect. 2.3 we calculate the stellar winds from main sequence (MS) and evolved stars, and characterize the impact of mass-loading from stars onto the radio AGN population (Sect. 2.2).} \addedit{In Sect.~\ref{sec:meth_radioagn} we characterize the radio AGN population and then calculate the stellar winds from main sequence (MS) and evolved stars in Sect.~\ref{sec:meth_ml}. We apply the model to our AGN population in Sect.~\ref{sec:results}, and in} Sect.~\ref{sec:disc} we examine the average impact of mass-loaded jets on galaxies, calculating a heating rate for galaxies of different masses and ages, which is needed for galaxy formation models. We finally assess the role of stochasticity from asymptotic giant branch (AGB) stars in deceleration of AGN jets in low mass galaxies (Sect.~\ref{sec:disc_stoch}).

\section{Methods}\label{sec:methods}
\subsection{Jet deceleration}\label{sec:meth_jet}

We treat mass-loading as a secular deceleration process. For the purposes of this paper, we define the criterion for jet deceleration as the Lorentz factor and location along the jet when half of its momentum is in the swept-up stellar material (that is, equal share of momentum flux in swept up mass and initial jet plasma.) In Sec.~\ref{sec:disc_stoch}, we discuss whether this assumption is valid for evolved star \addedit{mass-loss} in low-mass dwarf galaxies. Appendix~\ref{sec:app_thresh} also discusses two alternative thresholds and and shows that they do not significantly affect the results (see Fig.~\ref{fig:thresholdcomparison}).

\addedit{We consider a relativistic jet with constant total momentum flux across the jet's cross-sectional area. Note that here and throughout this section, we assume the ultra-relativistic limit, or $\beta\equiv v/c=1$. In what follows, we briefly discuss the condition for mass loading to slow down the jet that will be used throughout the rest of the paper, while appendix~\ref{sec:app_deriv} steps through this derivation in more detail. The momentum of the jet is the sum of the original jet plasma's momentum, $\dot{\Pi}_{\rm init} = P_j/c$ and the mass loaded material, $\dot{\Pi}_*$,
\begin{equation}
    \dot{\Pi}_{\rm tot} \equiv \dot{\Pi}_j + \dot{\Pi}_* = \frac{P_j}{c}\frac{\Gamma}{\Gamma_j} + \dot{\Pi}_*,
\end{equation}
where $\Gamma_j$ is the original bulk Lorentz factor of the jet, $\Gamma$ is the instantaneous Lorentz factor, and $P_j$ is the jet kinetic power. We can express the momentum flux of the swept up material as
\begin{equation}
    \dot{\Pi}_* = \Gamma c \int_0^{\dot{M}_j} \Gamma(\dot{M}_j')~ d\dot{M}_j'.
\end{equation}
The factor of $\Gamma$ inside the integral accounts for the internal motion at the time of mass-loading, which will decrease as more mass becomes loaded.} Due to passing through a shock as it is accelerated to the jet velocity, \addedit{the mass-loaded material} must have the same average internal (random) Lorentz factor $\langle \gamma \rangle = \Gamma$ as the bulk flow \addedit{at the time of entrainment} \citep[see also studies of entrainment in Gamma-Ray Bursts, e.g.,][]{Xie2012}. 

\addedit{By assuming that $\dot{\Pi}_{\rm tot}$ remains constant as the jet propagates, we can find the mass-loading rate such that a half of a jet's momentum is from entrained material:}
\begin{equation}
    \dot{M}_{j,\rm crit} \equiv \frac{3}{2}\frac{P_{j}}{c^{2}\Gamma_{j}^{2}}.
\end{equation}

Assuming a conical (ballistic), two-sided jet with a small opening angle $\theta$\removeedit{(an assumption that we discuss further in Sect. 4.2)}, we can relate the MLR in the jet ($\dot{M}_j$) to the MLR in the entire galaxy ($\dot{M}$) by
\begin{equation}
    \dot{M}_j = \frac{1}{2}\theta^2\dot{M} ,
    \label{eq:meth_frac}
\end{equation}
yielding a critical power
\begin{equation}
    P_\mathrm{crit} = \frac{1}{3} \theta^2 \Gamma_j^2 c^2 \dot{M}.
    \label{eq:P_crit}
\end{equation}

The critical jet power depends \addedit{linearly} on the galaxy-wide stellar mass loss rate as written in eq.~\ref{eq:P_crit}. Jets with higher Lorentz factors at a given power are slowed more easily due to the greater shock-acceleration entrained material must undergo into the jet rest frame. This expression does not \addedit{allow the} jet opening angles and stellar mass-loss rates \addedit{to} change as a function of distance from the center of the galaxy. In particular, we do not allow the jet to flare. \addedit{Recent works have found that a large fraction of local jets are conical on galactic scales \citep{Kovalev2020, Boccardi2021, Prameswari2025}, supporting our ballistic model.} 

Eq.~\eqref{eq:P_crit} resembles the standard relation from \cite{hubbard_active_2006}, where a jet is decelerated by mass-loading if $P_j < \Gamma_j \dot{M_T} c^2$. We include an additional factor of $\Gamma_j$ due to the additional acceleration of entrained material within the frame of the jet, as discussed above. (Note that their $\dot{M_T}$ refers only to mass loaded directly into the jet, and so corresponds to $\frac{1}{2} \theta^2 \dot{M}$, as we write $\dot{M}$ in terms of the mass-loss throughout the entire galaxy.) 

In order to remove the galaxy and black hole mass dependence from eq.~\eqref{eq:P_crit}, we can re-write it in terms of the Eddington ratio of the jet kinetic power 
\begin{equation}
    \lambda \equiv P_j/L_\mathrm{edd},
    \label{eq:edd_frac}
\end{equation}
where
\begin{equation}
    L_\mathrm{edd} = \frac{4\pi G M_\mathrm{bh} m_\mathrm{p} c}{\sigma_\mathrm{T}} \approx 1.3 \times 10^{38} \left( \frac{M_\mathrm{bh}}{\Msun} \right) ~\mathrm{erg ~s^{-1}}.
    \label{eq:L_edd_def}
\end{equation}

Given a linear black hole mass-stellar mass relation, we can completely remove the mass dependence from the critical Eddington ratio. Various black hole-stellar mass relations are presented in the literature \citep[e.g.,][]{Haring2004,Kormendy2013, McConnell2013,Reines2015}. We follow the relation found by \cite{Marconi2003}, $M_\mathrm{bh} = 0.002 M_*$, in order to be consistent with the radio AGN jet power distribution introduced in Sect.~\ref{sec:meth_radioagn} from \cite{Igo2024}. Making both of those changes, we can rewrite the critical jet Eddington ratio as:

\begin{equation}
    \lambda_\mathrm{crit} = \frac{ \theta^2 \Gamma_j ^ 2 c^2}{7.8 \times 10^{35}~\mathrm{erg~s^{-1}}}  \dot{m}_{\rm stellar},
    \label{eq:lambda_crit}
\end{equation}
\addedit{where $\dot{m}_{\rm stellar} \equiv {\dot{M}}/{M_*}$, the mass-loss rate per stellar mass. As we show in Figure~\ref{fig:mlr_galfrac}, $\dot{m}_{\rm stellar}$ is independent of $M_*$. Hence,} this expression allows us to calculate the maximum jet Eddington fraction that can be stopped by a certain entrainment rate, \addedit{depending only on jet properties, and} independent of the stellar mass of a host galaxy. \removeedit{, and enables us to work consistently with the radio AGN incidence data outlined in the following section.}

\subsection{Radio AGN incidence}\label{sec:meth_radioagn}
To study the role of mass-loading in the local jet and galaxy populations, we must characterize how jet powers are distributed as functions of their host properties. Radio AGN incidence evolves through the parameter space of jet Eddington ratio, stellar mass, \addedit{color}, and redshift \citep[see e.g.,][and references within]{Delvecchio2022, Magliocchetti2022, Zhu2023}. Here, we focus on radio AGN in the local Universe, using data from \cite{Igo2024} and \cite{Igo2025}.

Using the LoTSS survey, \cite{Igo2024} studied the incidence of radio AGN, characterized by \addedit{jet} Eddington ratio and morphology (complex vs compact), and matched radio AGN with host galaxies from the GAMA09 sample to study connections with host mass and environment. When binned by stellar mass and redshift, the study found that more massive galaxies were more likely to host radio AGN across their entire $\lambda$ range, consistent with literature \citep[e.g.,][]{Best2005, Hickox2009, Ching2017, Williams2018, Sabater2019, Mingo2022}. Here, we represent the radio AGN incidence as $\eta$, referring to the fraction of galaxies in each mass bin hosting a radio-loud AGN per unit $\log\lambda$, with $\lambda$ defined as in eq.~\eqref{eq:edd_frac}.

There is significant uncertainty in the derivation of the jet kinetic power, which usually involves a conversion from a luminosity measurement in a given frequency band. A variety of power law $P-L_\mathrm{R}$ relations have been proposed \citep[e.g.,][]{Best2006, Heckman2014, Hardcastle2018}. Other relations account for other factors, such as radio spectral index \citep{heinz_non-linear_2003, Merloni2003}, or proton content \citep{Croston2018}. We do not arbitrate between these different relations here, but they are discussed in greater detail in \cite{Igo2024} Sect.~6.2. Also, see \cite{Igo2024} Sect.~3.2.4 for further details about sample selection methods, such that we are working with a largely mass- and luminosity- complete sample.

\begin{figure}
    \centering
    \includegraphics[width=\linewidth]{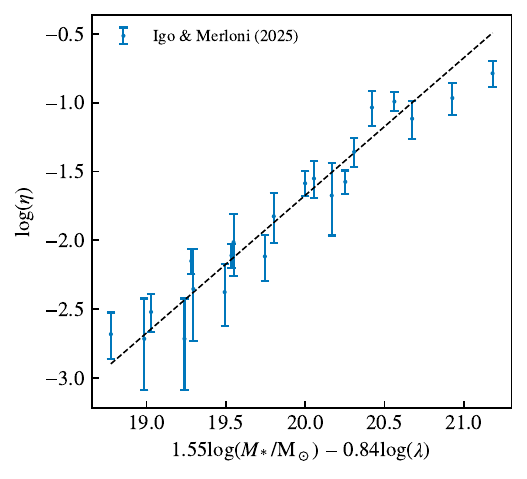}
    \caption{Edge-on view of our planar fit (dashed line) to log incidence $\eta$ (normalized per dex in jet Eddington ratio) of radio AGN, as a function of log($\lambda$) and stellar mass.}
    \label{fig:eddfrac_plane_fit}
\end{figure}

With normalized per dex data of this sample (combining complex and compact sources) from \cite{Igo2025} at $z \leq 0.285$, we use orthogonal distance regression to fit a plane to the log-spaced data in stellar mass and jet Eddington ratio, using the bin widths as estimates of our uncertainties. This plane is shown side-on in Figure~\ref{fig:eddfrac_plane_fit}, taking the form 

\begin{equation}
    \log \eta = a \log M_* + b\log \lambda + c,
    \label{eq:radioagn_incidence_fit}
\end{equation}
with $a = 1.55 \pm 0.160$, $b = -0.845 \pm 0.074$, and $c = -21.7 \pm 1.79$.

Fig.~\ref{fig:eddfrac_plane_fit} shows that a plane is a reasonable fit to the data, though more complex functions can match the data more exactly \citep[see further discussion in][]{Igo2025}. \addedit{In Sect.~\ref{sec:res_frac}, we introduce two approaches to using this incidence data, the first (our `baseline sample') limited to the $\lambda$ range of \cite{Igo2024} and the second with the planar fit extrapolated to account for a potentially unobserved population.}\removeedit{Our assumption of a planar fit becomes obviously unphysical at small Eddington fractions, as the cumulative incidence of radio AGN for a given galaxy population cannot exceed 1. We discuss this choice further in Sect.~4/4, however, our critical Eddington fractions are typically $\lambda_{\rm crit} > 10^{-3}$ and the shape of the incidence function below $\lambda_{\rm crit}$ is not important to our calculations.}

As the errors in the best fit plane are correlated, we use a Monte Carlo approach to constrain the uncertainty in the planar fits. For the calculations through the rest of the analysis, we use 1000 bootstraps of radio AGN incidence, drawing from a Gaussian distribution in the spread in the logspace bin widths in stellar mass and $\lambda$, as well as the uncertainty on AGN incidence.

\subsection{Mass loss from stellar winds}\label{sec:meth_ml}
In this section, we develop mass loss models for a variety of stellar winds, in an effort to characterize under what conditions different masses, and evolutionary phases, of stars contribute most greatly to the mass available for entrainment. We use observations of X-ray luminosity as a proxy for MLRs among low mass MS stars, and simulation results from the MIST \citep[MESA Isochrones and Stellar Tracks;][]{Paxton2011, Paxton2013, Paxton2015, Choi2016, Dotter2016} evolutionary pathways for high mass MS stars as well as evolved stars. MIST does not accurately model coronal mass loss for main sequence stars, necessitating our use of X-ray luminosity as a proxy to derive MLR estimates.

\subsubsection{Low-mass main sequence stars}\label{sec:ml_ms}
Developing mass loss models for low mass main sequence stars is challenging, as direct measurements of thermally-driven, hot coronal stellar winds are limited to our own Sun. As a result, our knowledge of low-mass stellar winds is largely derived from the solar wind. Although the Sun is thought to be a fairly representative solar-type star \citep{gustafsson_is_1998}, whether its mass loss rate properties can be extrapolated to other types of stars is poorly understood. \removeedit{Though transient absorption features (e.g., H$\alpha$, Ca, and K lines) have been used to constrain mass loss rates for fast rotating stars,}Most indirect measurements of low mass MS mass loss rates until now have been done through H\,I Lyman-$\alpha$ absorption in the astrosphere \citep{wood_new_2021}. Even so, this method has only yielded on the order of 20 detections. 

Stellar winds are associated with the star's magnetic activity through coronal heating. For this reason, one approach to calculating mass loss rates is to use the X-ray luminosity ($L_X$) as a proxy for stellar activity. \addedit{See Appendix~\ref{sec:appendixB} for a brief discussion of the literature surrounding X-ray luminosities and mass-loss. }

In the absence of an established approach to estimating coronal MLR, we will follow the simple parametric assumption that $\dot{M} \propto L_X$, normalized to the solar wind mass loss rate, $\dot{M}_{\odot} \approx 2 \times 10^{-14} ~\mathrm{\Msun~yr^{-1}}$ \citep{cohen_independency_2011}, and X-ray luminosity $L_{\rm X, \odot} \approx 10^{27}~ \mathrm{erg~s^{-1}}$ \removeedit{(see } \citep{peres_sun_2000} \removeedit{for upper and lower limits, log($L_X$) = 26.43 and 27.67 respectively, from variation throughout the solar cycle)}.

\begin{figure}
    \centering
    \includegraphics[width=\linewidth]{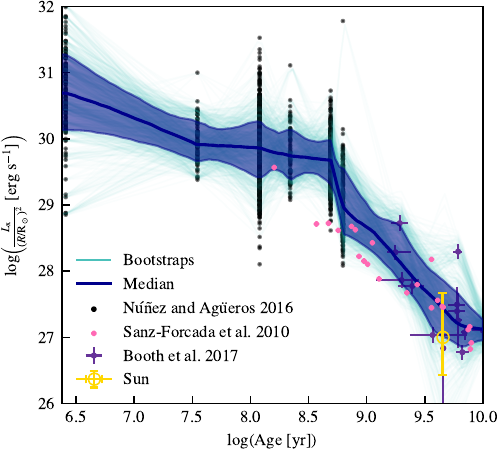}
    \caption{X-ray flux through the stellar surface as a function of stellar age. The black points show young open cluster data from \cite{Nunez2016}, and the purple points show stars measured by \cite{Booth2017}. Thin teal lines display the bootstraps as described in the text. The dark blue line and shaded region show median and 1$\sigma$ distribution of the bootstrapped data. The Sun is shown in yellow, at an age of 4.5 billion years, and the pink points show data from \cite{SanzForcada2010} for reference.}
    \label{fig:bootstrap_Lx}
\end{figure}

We construct samples of stellar X-ray luminosity with open cluster data from \cite{Nunez2016} and astroseismic targets from \cite{Booth2017}, all of which include G-, K-, and M-stars. \addedit{See Sect.~\ref{sec:appb_dataprep} for further discussion of how we prepare the data.}

Figure~\ref{fig:bootstrap_Lx} shows stellar X-ray flux as a function of stellar age\removeedit{. These}\addedit{, with }bootstrapped profiles shown in teal \addedit{(see Sect.~\ref{sec:appb_dataprep})}. The median and 1-$\sigma$ envelope are drawn and shaded in dark blue, respectively. (All other uncertainties in this work are shown at 1-$\sigma$, unless otherwise stated.)

In Fig.~\ref{fig:bootstrap_Lx}, we also overplot X-ray luminosity and age data for stars from \cite{SanzForcada2010}\removeedit{, with radii derived from the stellar spectral types converted to stellar radius using the relations of Pecaut and Mamajek 2013. Only stars with published ages, X-ray luminosities, and a spectral type are shown}. Though we do not include this data for the bootstrapping and fitting due to the lack of available uncertainties, it is generally consistent with our relation. The Sun is also shown for reference, and is consistent with the low-$L_X$ side of the relationship as expected for a low-activity star \citep{peres_sun_2000}. 

Converting the X-ray luminosities shown in Fig.~\ref{fig:bootstrap_Lx} to mass loss rates requires certain assumptions, \addedit{as discussed in Sect.~\ref{sec:appb_lx_mdot},} due to the lack of observational data on MLRs for stars other than the Sun. In particular, we assume that $\dot{M}$ is proportional to $L_X$ for individual stars, normalized by $\dot{M}_\odot$ and $L_{X\odot}$.\footnote{To account for the radius dependence in the flux calculation, we assume $R/R_\odot \propto (M/M_\odot)^{0.8}$.} Then, we can write the mass loss rate of the stellar population (divided by galaxy stellar mass) as a function of stellar population age, $\tau$:
\begin{equation}
    {\dot{m}_{\rm MS}}(\tau) = \frac{\dot{M}_\odot}{L_{X\odot}} \frac{L_X}{(R/\rm R_\odot)^2}(\tau) \int_{0.08 \Msun}^{M_\mathrm{max}(\tau)} \phi_\mathrm{IMF} ~\left(\frac{M}{\rm M_\odot} \right)^{1.6}~ dM,
    \label{eq:mlr}
\end{equation}
where $\frac{L_X}{(R/\rm R_\odot)^2}(\tau)$ is our relationship from Fig.~\ref{fig:bootstrap_Lx}, and $\phi_\mathrm{IMF}$ is the initial mass function (IMF) of the stellar population. Here, we use the Kroupa IMF for our assumed stellar population \citep{kroupa_variation_2001}. The choice of IMF is not particularly crucial; use of a \cite{chabrier_galactic_2003} IMF instead has a negligible effect on the results. The mass-weighted IMF is normalized only from 0.08\,M$_\odot$ to the main sequence turn-off mass $M_\mathrm{max}(\tau)$, which also does not significantly affect the results. 

The mass loss rate through the entire galaxy (scaled by galaxy stellar mass) from various stellar types is shown in Figure~\ref{fig:mlr_galfrac}, with the blue curve representing the mass-loss contributions of low mass main sequence stars calculated with eq.~\eqref{eq:mlr}. The shape of the curve is, as expected, generally proportional to that of the stellar X-ray luminosity, with slight adjustments at $\tau{\sim}10$~Gyr due to the turn-off mass dropping into the range of G-type stars.

\removeedit{Throughout, we ignore the contribution of main sequence A- and F-type stars, whose winds are poorly understood. The lack of a convective envelope in such stars would seem to rule out coronal activity and thermal winds. If radiative winds do exist for such stars (e.g., Landstreet1998) their MLR is thought to be extremely small, e.g., $\dot{M}\sim10^{-16}~\Msun~ \rm yr^{-1}$ (Babel1995). A-stars are relatively uncommon compared to their lower-mass MS counterparts, and combined with these small MLRs, the exclusion of mass loss from A- and F-stars should minimally affect our findings.}

\subsubsection{Evolved stars}\label{sec:ml_evolved}

\begin{figure}
    \centering
    \includegraphics[width=\linewidth]{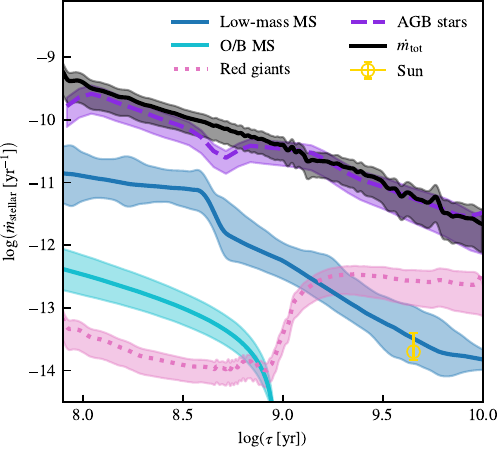}
    \caption{Mass loss rate (scaled by galaxy stellar mass) as a function of stellar population age for a variety of stellar types. Low-mass main sequence stars (blue line) account for G-, K-, and M-type stars, while the cyan line shows mass loss from main sequence O- and B-type stars. The contribution of evolved stars are shown, including red giants (pink dotted), and AGB stars (purple dashed). Finally, the black curve shows mass loss leaving the IMF.}
    \label{fig:mlr_galfrac}
\end{figure}

Evolved stars shed mass at rates far greater than their low mass MS progenitors, though they are fewer in number. To model their contribution, we use evolutionary tracks from MIST, using the standard rotation and solar metallicity tracks (our results do not vary greatly with reasonable variations of these parameters). We interpolate the data to calculate\removeedit{when stars of given masses leave the main sequence, the red giant branch, AGB, etc. Their masses at each time step are also available. The} \addedit{average} mass loss rates over the \removeedit{red giant and AGB periods are averaged over the}respective lifetimes of the \addedit{red giant and AGB} evolutionary phases \citep[washing out, e.g., the large oscillations in MLR during the thermally pulsing AGB phase, see][and references therein. If there are a large number of AGB stars in the jet, then this temporal average is valid.]{Hofner2018} Then the contribution to the entire galaxy (per stellar mass) is calculated as
\begin{equation}
    {\dot{m}_{\rm evolved}}(\tau)= \int_{M_\mathrm{min, ~ev}(\tau)} ^ {M_\mathrm{max,~ev}(\tau)} \dot{M}_{\rm ev}(M) ~\phi_\mathrm{IMF} ~dM,
    \label{eq:evolved_star}
\end{equation}
where $\dot{M}_{\rm ev}(M)$ is the average AGB or red giant MLR as a function of MS stellar mass, and $M_\mathrm{min}$ and $M_\mathrm{max}$ bound the stellar mass ranges residing in each evolutionary phase at a given time. The red giant and AGB phase \addedit{wind loss rates} are shown by dotted pink and dashed purple lines, respectively, in Fig.~\ref{fig:mlr_galfrac}. Mass loss from red giant stars exceeds main sequence stars only in old stellar populations $\tau \gtrsim10^{9.5}$~yrs. AGB stars, though fewer, contribute significantly more than red giants to the total stellar mass-loss on a galactic scale. In Sect.~\ref{sec:disc_stoch} and \ref{sec:disc_cav_incidence}, we consider the implications of having a very small number of AGB stars in a galaxy on this methodology.

\subsubsection{High-mass main sequence stars}\label{sec:ml_hmms}
For \addedit{completeness}\removeedit{comparison}, we also calculate the contribution of high-mass O- and B-type stellar winds to the galaxy. Such winds are radiatively driven, with individual mass-loss rates reaching up to $10^{-5}~\Msun~\mathrm{yr^{-1}}$ \citep[for more discussion of radiative winds, see][and references within]{Smith2014}. For our estimates, we again utilize MIST evolutionary tracks and average the mass lost over the stellar main sequence lifetime, integrating over the IMF following eq.~\eqref{eq:evolved_star}. The contribution of these stars rapidly decreases as O-type stars leave the main sequence, falling below the low-mass stellar population by $\tau\sim 10^{7.2}$\,yrs and quickly becoming negligible, as shown in Fig.~\ref{fig:mlr_galfrac}. We thus ignore high-mass MS stars in the remainder of our analysis.

\addedit{Throughout, we ignore the contribution of main sequence A- and F-type stars, whose winds are poorly understood. The lack of a convective envelope in such stars would seem to rule out coronal activity and thermal winds. If radiative winds do exist for such stars \citep[e.g.,][]{Landstreet1998} their MLR is thought to be extremely small, $\dot{M}\sim10^{-16}~\Msun~ \rm yr^{-1}$ \citep[][]{Babel1995}. A-stars are relatively uncommon compared to their lower-mass MS counterparts, and combined with these small MLRs, the exclusion of mass loss from A- and F-stars should minimally affect our findings.}

\subsubsection{Integral constraints on stellar mass loss calculated from the evolution of the stellar mass function}\label{sec:ml_imf}
The amount of mass in stars leaving the MS (via the IMF) provides an upper limit on the mass lost from stars, while neglecting entirely the details of the physical processes underlying that mass loss. 

We can calculate the \addedit{rate of} stellar mass leaving the main sequence as:
\begin{equation}
    {\dot{m}_{\rm IMF}}(\tau) = -\frac{\rm d}{\rm d\tau} \int_{0.08}^{M_\mathrm{max}(\tau)} \phi_\mathrm{IMF}~M~dM.
    \label{eq:Mdot_off_MS}
\end{equation}
We can then estimate the mass in collapsed stellar remnants (i.e., white dwarfs for stars in the mass ranges we are interested in here) as
\begin{equation}
    {\dot{m}_{\rm WD}}(\tau) = \frac{\rm d}{\rm d\tau} \int_{M_\mathrm{WD}(\tau)}^{8} \phi_\mathrm{IMF}~M_\mathrm{WD}~dM,
    \label{eq:Mdot_WD}
\end{equation}
using a white dwarf mass function derived from MIST. The difference between these two rates, \addedit{$\dot{m}_\mathrm{IMF}(\tau) - \dot{m}_\mathrm{WD}(\tau)$,} must be released through mass loss into the surrounding interstellar medium.\removeedit{:
\begin{equation}
    {\dot{m}_{\rm tot}(\tau)=\dot{m}_\mathrm{IMF}(\tau) - \dot{m}_\mathrm{WD}(\tau).}
    \label{eq:Mdot_tot}
\end{equation}}

This analytical approach, shown as a black line in Fig.~\ref{fig:mlr_galfrac}, agrees to within ${\sim}10\%$ with a mass loss rate calculated by summing all of the other contributing mass losses shown in Fig.~\ref{fig:mlr_galfrac}, along with (uncertain) post-AGB mass loss derived from MIST. Thus we adopt this analytical approach\removeedit{of eq. 14} throughout for an estimate of total mass loss in a galaxy as a function of its stellar population age.

\addedit{Note that we ignore mass injection from supernovae. Energy injection from supernovae can significantly affect galaxies, and jet-supernova interactions may be an important particle acceleration site \citep[][]{torresalba2019, boschramon2023}. However, the global mass injection rate is significantly lower than that of stellar winds \citep[e.g., ${\sim}1\%$ the wind rate,][]{voit2015, boschramon2023}. Mass injection from supernovae in jets is also highly stochastic due to the low supernova rate \citep[see e.g.,][]{Mohapatra2024}, and thus would not be well-captured by our model.}

\section{Results}\label{sec:results}
We find that, over the stellar population ages considered ($\tau \gtrsim 10^{8}~{\rm yrs}$) stellar mass-loss in galaxies is dominated by AGB stars at all times, as shown in Fig.~\ref{fig:mlr_galfrac}. Still, it is worth noting that around 1~Gyr, mass loss from low mass main sequence stars comprises ${\sim}10\%$ of total mass-loss (see Sect.~\ref{sec:disc_stoch} for further discussion on the role of main sequence stars). The relative contribution of red giants increases steeply for 1\,Gyr stellar populations, which is expected as the MIST modeling for red giant branch mass loss increases at that stellar turn-off age corresponding to ${\sim} 2\,\Msun$ stars. Mass loss on the red giant branch \removeedit{(RGB) }is thought to be negligible for higher mass stars \citep{Hofner2018}, consistent with our finding for the population. The MIST AGB star mass loss modeling is also consistent with the observations of \cite{Cummings2016, Hofner2018}, where mass loss during the AGB phase increases from approximately 25\% to 80\% of initial mass from 1\,$\Msun$ to above $4\,\Msun$.


\begin{figure}
    \centering
    \includegraphics[width=\linewidth]{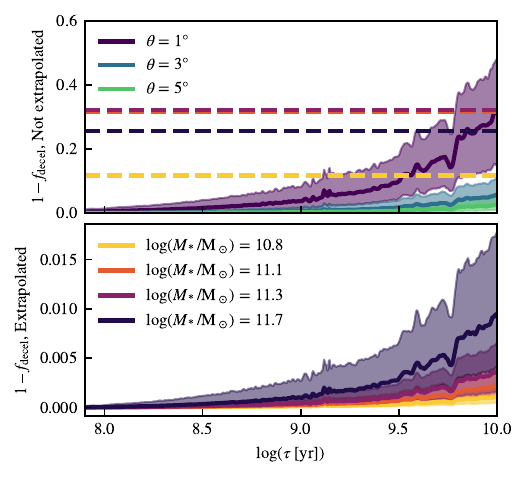}
    \caption{Fraction of jets not decelerated by mass loading. The top panel shows the escape fraction of jets \addedit{above the \cite{Igo2024} jet Eddington fraction range, for a range of opening angles. The horizontal dashed lines represent the FR-II fraction found in \cite{Igo2024} for varying stellar mass bins, though note that these measurements have large errors \citep[see Table 2 in][]{Igo2024}. In the bottom panel, we use the extrapolated sample,} for a ${\sim}2$~dex range of stellar masses. The bottom panel assumes $\theta=3\degree$, and both panels adopt $\Gamma_j=3$.}
    \label{fig:f_stopped}
\end{figure}

Relative to other studies of stellar mass-loading in jets, generally in red ellipticals, our \addedit{specific} mass loss rates of ${\dot{m}_{\rm stellar}}\lesssim10^{-11}\,\rm yr^{-1}$ are higher than those used by the early numerical models of \cite{Komissarov1994} and \cite{bowman_deceleration_1996}, but within the range utilized by \cite{angles-castillo_deceleration_2021}, \addedit{and those predicted by recent works that consider the specific origin of mass-loss \citep[e.g.,][who find $\dot{m}_{\rm stellar} = \rm few \times10^{-12} ~\rm yr^{-1}$ for Cen A]{Wykes2019}.} Thus, we expect to find an enhanced effect of mass-loading \removeedit{(i.e. deceleration of ${\sim}1$~dex more powerful jets)} relative to those early studies.

\subsection{Main sequence mass-loss}\label{sec:res_ms}

Reliable age dating of stars older than ${\sim}1~$\,Gyr is at present generally limited to a small number of asteroseismic targets. That region of our X-ray luminosity profile in Fig.~\ref{fig:bootstrap_Lx} is thus based on a limited amount of stellar data from \cite{Booth2017}. Upcoming surveys aim to measure astroseismic stellar ages, i.e. through high-precision radial velocity data or the PLATO survey \citep[PLAnetary Transits and Oscillations of stars, see e.g.,][]{Betrisey2023, Rauer2025}. As such programs uncover a larger census of older stars, the accuracy of X-ray profiles such as this one will become better constrained. 

It is encouraging that our X-ray surface flux results are in line with stellar data that we do not use to fit, such as the catalogs from \cite{SanzForcada2010, SanzForcada2011} which we exclude due to the lack of published uncertainties. We can also compare to theoretical works: \cite{Johnstone2021} created models for stellar X-ray evolution of F-, G-, K-, and M-type stars. For their 5\,Gyr sample ($\log[\rm Age] \sim9.7$), they found X-ray luminosities between $10^{27}$ and $10^{28}\,\mathrm{erg~s^{-1}}$, with little dependence on rotation speed, and with lower mass stars accounting for the lower X-ray emission. Accounting for our normalization by stellar surface area (which will slightly boost $L_X$ for the lower mass stars), their predictions are consistent with our findings in Fig.~\ref{fig:bootstrap_Lx}.


\begin{figure}
    \centering
    \includegraphics[width=\linewidth]{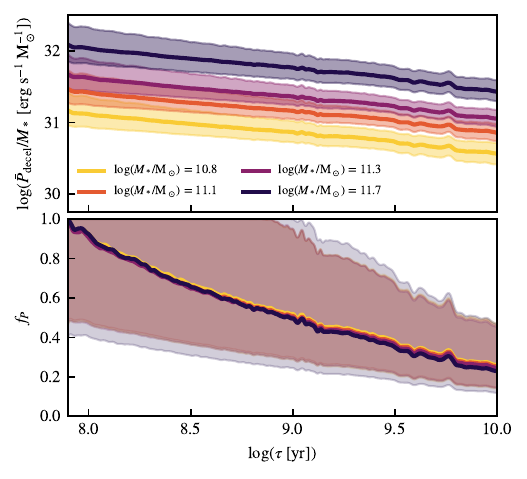}
    \caption{Feedback power in decelerated jets. The top panel shows the mean kinetic power decelerated per stellar mass (i.e. specific feedback power) as a function of stellar population age, for selected stellar masses. The bottom panel shows the fraction of kinetic power in jets that is affected (i.e. the top panel divided by the total mean jet power for a given stellar mass). Throughout we use our fiducial model of $\theta=3\degree$, and $\Gamma_j=3$.}
    \label{fig:power_plots}
\end{figure}

We observe a break in the X-ray surface flux at about $\log(\rm Age)\sim8.5$, after which $L_X$ decreases more rapidly with time. This break is also seen in the M-dwarf sample of \cite{Engle2024}, at ages ranging between $\log(\rm Age)\sim8.6-9.3$, depending on the type of M-dwarf. They propose that the break is due to the competing effects of rotation and convection. \cite{wood_new_2005} also identify a break in this relation at an age corresponding to 0.7\,Gyr (log[Age] = 8.85), at the same position with the break we identify in Fig.~\ref{fig:bootstrap_Lx}. Recent studies also measure a break in $L_X$ as a function of Rossby number in M-dwarfs \citep{Shoda2021,Landin2023, Shan2024}. \citep[The Rossby number is thought to relate to stellar age by $Ro\propto t^{0.5}$, e.g.,][]{Vardavas2005}. Expanding our stellar sample would improve the statistics on our X-ray surface flux-age relation, but as is, it reproduces the main features seen throughout the literature. Given the uncertainties in our estimates, the presented heuristic estimates of MLRs are therefore sufficient for our purposes.

\subsection{Likelihood of jet deceleration and jet feedback}\label{sec:res_frac}

Any individual galaxy may undergo phases of varying jet powers and changes in its stellar population, therefore, we seek to understand the ensemble properties of jets in different types of galaxies and, further, the amount of kinetic feedback enabled by stellar mass loading. Given the radio AGN incidence relation shown in Fig.~\ref{fig:eddfrac_plane_fit}, alongside the MLRs displayed in Fig.~\ref{fig:mlr_galfrac}, we can calculate in a statistical way how jet populations respond to stellar mass-loading. First, we calculate the fraction of jets (by number, rather than power) that escape undecelerated from their hosts. We then examine the fraction of radio AGN kinetic power in jets decelerated by their host galaxy, and thus available for kinetic feedback in the host galaxy. 

We can calculate the likelihood that stellar mass-loading decelerates a jet within its host galaxy. To do so, we simply transform the radio AGN incidence shown in Fig.~\ref{fig:eddfrac_plane_fit} into a cumulative incidence function, and find the fraction of jets with $\lambda < \lambda_\mathrm{crit}$, labeled as $f_{\rm decel}$. 

\addedit{We treat the cumulative incidence in two ways: (i) our baseline sample, fixing the minimum jet power at the lowest end of the \cite{Igo2024} incidence function ($\lambda_{\rm min} = 10^{-4}$). This method allows us to apply our model to the exact sample used by \cite{Igo2024} and compare our predictions to their findings. We also (ii) extrapolate the incidence function such that the cumulative incidence reaches 1 for each mass bin (i.e., assuming every galaxy is switched on at some low level). $\lambda_{\rm min}$ is then the jet Eddington fraction where the cumulative radio AGN incidence reaches 1. This extrapolation allows us to make predictions about the inferred complete jet population, including sources that go unobserved due to their size or faintness. \cite{Sabater2019} found that all galaxies in their sample with $M_*>10^{11}~\Msun$ had radio AGN emission at some level, suggesting that our extrapolation may be useful, though it may not hold for lower-mass galaxies.}

The results of that calculation are shown in Figure~\ref{fig:f_stopped}. Note that we in fact plot $1-f_{\rm decel}$ for legibility -- the fraction of jets that are not decelerated. The top panel shows the escape fraction as a function of stellar population age, for a range of opening angles $\theta$, \addedit{calculated using the baseline sample. As stellar populations age and their MLRs decrease, a larger fraction of jets are able to escape undecelerated. Similarly, as $\theta$ decreases, so does the number of stars in the jet, and more jets are able to escape. For an old stellar population and our fiducial case ($\theta=3\degree,~\Gamma_j=3$), we estimate that ${\sim}10\%$ of jets escape, though a small decrease in opening angle has a strong impact on the deceleration: if $\theta=1\degree$, ${\sim}35\%$ of jets are not decelerated. For old stellar populations, our escape fraction is in line with the predicted FR-II fraction from the \cite{Igo2024} sample. Using the baseline sample and our planar incidence fit, the escape fraction is not a function of stellar mass.}

\addedit{The bottom panel of Fig.~\ref{fig:f_stopped} shows again the escape fraction, but calculated using the extrapolated sample. The extrapolation produces an additional inferred population of low $\lambda$ jets, which are efficiently decelerated according to our mass-loading model. We find that mass-loading from \addedit{the winds of} a stellar population can effectively decelerate ${\gtrsim}99\%$ of these jets across all stellar mass ranges \addedit{and stellar population ages}. Using the extrapolation, more massive galaxies are more likely to host an AGN jet that is not decelerated, as opposed to the mass independence of the escape fraction in the baseline sample. This mass dependence arises because more massive galaxies are more likely to host radio AGN in the \citet{Igo2024} range, so the low-mass galaxies must reach a range of lower $\lambda$ in order to have a cumulative incidence of 1. In this sense, the mass-dependence may be an unreliable artifact of the extrapolation. A more complex model of cumulative radio AGN incidence than our simply extrapolation, like the one presented by \cite{Sabater2019}, might affect or reduce the mass-dependence.}

\removeedit{Additionally, more massive galaxies are more likely to host an AGN jet that is not decelerated. Such a trend may seem counterintuitive, because the MLR of a galaxy scales linearly with stellar mass. However, our fit of radio AGN incidence scales superlinearly with $M_*^{1.55}$ (see the fit discussed in Sect. , as well as Igo+24.) This increase in jet incidence overcomes the linear increase in MLR, such that jets in more massive galaxies are more likely to be powerful enough to escape their hosts.}

The amount of kinetic power globally available for \addedit{direct} radio mode feedback inside galaxies, \addedit{as opposed to larger-scale preventative feedback,} depends on the efficient deceleration of jets within their host galaxies. To this end, we\removeedit{must} calculate the ensemble average of decelerated kinetic power, which is shown in the top panel of Figure~\ref{fig:power_plots} as a specific quantity for selected stellar masses. As expected, more massive galaxies can slow down more powerful jets, but $P_{\rm crit}$ evolves with stellar population age, showing a decrease of ${\lesssim}0.5$~dex in power over 2\,dex in age.

We find the mean jet power decelerated for a given host stellar mass with
\begin{equation}
    \bar{P}_\mathrm{decel}(M_*,\tau) = \int_{P_\mathrm{min}(M_*,\tau)}^{P_\mathrm{max}(M_*,\tau)} P ~\eta~ dP,
    \label{eq:P_calc}
\end{equation}
where $P_\mathrm{min}$ is the minimum jet power based on our fitted plane of radio AGN incidence, and $P_\mathrm{max} = P_\mathrm{crit}$, which we can write as $\lambda_\mathrm{crit}L_\mathrm{edd}$ for a given mass from eq.~\eqref{eq:edd_frac}. As before, $\eta(\lambda,M_*)$ is the incidence per $\log \lambda$, transformed into incidence per power for a given stellar mass.

\addedit{For the baseline sample, $P_{\rm min}$ is simply equivalent to $\lambda_{\rm min} L_{\rm edd}$. With the extrapolated sample, $P_{\rm min}$ is instead the power where the cumulative radio AGN incidence reaches 1. Note that in Fig.~\ref{fig:power_plots} we show only results for the extrapolated sample, as the two methods vary by $\lesssim5\%$ at all times.} 

From the mean power, we are able to calculate the fraction of stopped jet power, $f_P$, shown in the bottom panel of Fig. 5, by dividing $\bar{P}_{\rm decel}$ by the mean power in all jets, whether decelerated or not [eq.~\eqref{eq:P_calc}, setting $P_\mathrm{max}=L_\mathrm{edd}$]. \addedit{Interestingly, although the two samples yield significantly different jet escape fractions (${\sim}1$\,dex), $\bar{P}_{\rm decel}$ and $f_P$ are far less affected. The extrapolation adds a large number of lower-power jets, but $\bar{P}_{\rm decel}$ is weighted towards high-power sources. Therefore, calculations of the energy available for feedback in the ISM (e.g., in the discussions in Sect.~\ref{sec:disc_energetics}) are largely invariant to the choice of method.}

The host stellar mass dependence in this power fraction $f_P$ is greatly reduced compared to the fraction of jets slowed by number (Fig.~\ref{fig:f_stopped}), meaning that the \textit{fraction} of jet power available \addedit{for direct feedback (i.e.} radio mode feedback \addedit{ on galaxy scales} in the local Universe\addedit{)} is essentially a galaxy-mass-independent quantity.

However, $f_P$ does exhibit a significant dependence on stellar age due to the strong dependence of the MLR on stellar age. We find fairly large uncertainties in this power fraction due to the combination of uncertainty in the radio AGN incidence, mass loss rate, and IMF. Even so, we calculate that at all times, $\gtrsim25\%$ of jet kinetic power in low-$z$ jets is affected by stellar mass loading when evolved stars are considered, producing kinetic power that can then be coupled to the host to produce jet-mode feedback. \addedit{The amount of thermal energy the ISM receives from the local jet population will be further reduced from $f_P$ depending on the intrinsic coupling efficiency of the decelerated jet to its surroundings \citep[e.g.,][]{Mukherjee2016, Asahina2017, Rossi2024}.}

\begin{figure}
    \centering
    \includegraphics[width=\linewidth]{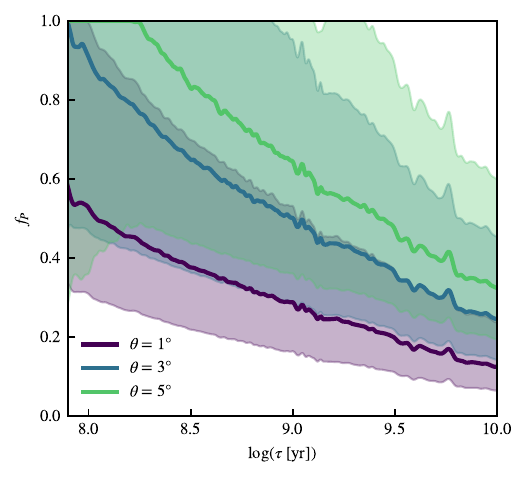}
    \caption{Fraction of power in the decelerated jet population for different opening angles, calculated for the extrapolated radio AGN incidence. We adopt $\Gamma_j=3$, and $M_* = 10^{11.3}\,\Msun$ (though $f_P$ is roughly mass-independent).}
    \label{fig:theta_variation}
\end{figure}

$f_P$ is well-described by a power law relation with $\tau$. For \addedit{the extrapolated sample}, $\Gamma_j =3,~\rm and~ \theta=3\degree$ (and over all masses), we find that,
\begin{equation}
    \log f_P = -0.31_{-0.15}^{+0.13} \log\tau + 2.49_{-1.04}^{+1.20},
\end{equation}
though the uncertainties are large. For a jet population with parameters $\theta,\Gamma_j$ close to our fiducial values, this relation can be used as a correction factor for jet-mode feedback energy deposition within host galaxies. Variations in $\Gamma_j ~\rm and ~\theta$ affect old stellar populations less strongly than for young stellar populations, within a factor of $\sim 2$ as shown in Fig.~\ref{fig:theta_variation}.

It is also important to note that Fig.~\ref{fig:power_plots} \removeedit{and 4} shows results only for our fiducial parameters $\theta=3\degree$ and $\Gamma_j=3$. From eq.~\eqref{eq:lambda_crit}, we found that $\lambda_{\rm crit}$ is proportional to MLR scaled by stellar mass. However, the opening angle $\theta$ and bulk Lorentz factor $\Gamma_j$ also affect $\lambda_{\rm crit}$ for a given mass loss rate. Although in some cases, e.g., Figs.~\ref{fig:mlr_galfrac} and \ref{fig:agb_stochasticity}, the results scale with the inverse square of $\theta$ and $\Gamma_j$, in most other cases the relation is modulated through the radio AGN incidence and is thus more complex. Fig.~\ref{fig:theta_variation} explores the effect that the opening angle has on \removeedit{the fractional power in jets affected by stellar mass-loading, }$f_P$. Larger opening angles increase the availability of jet kinetic power for feedback, and vice versa for smaller opening angles. For a $1\degree$ opening angle jet, we expect only $\sim$15\% of jet power available for feedback among old stellar populations, as opposed to $\sim$40\% for a $\theta=5\degree$ jet. To create a truly global $f_P$ function, integrating over $\theta$ and $\Gamma_j$ distributions would be necessary. As those distributions are not well-constrained, particularly that of the intrinsic opening angle, we leave that calculation for future work.

\section{Discussion}\label{sec:disc}
We have used a simple model to examine under what conditions jets are decelerated through stellar mass-loading. In this section, we discuss our findings in greater detail. In particular, we comment on jet contributions to global energetics (Sect.~\ref{sec:disc_energetics}) as well as our predictions for jet morphology (Sect.~\ref{sec:disc_morph}). We also find a threshold in jet power below which main sequence stars dominate the deceleration of jets and discuss the importance of AGB stochasticity in dwarf galaxies, independent of host mass.

The total MLR of a galaxy undergoes a strong time evolution, decreasing by ${\sim}$2~dex between populations of $\tau=10^8 - 10^{10}$ years. (Main sequence stars decrease even more drastically over the same age range, ${\sim}$3~dex.) The total mass loss rate evolves as $\tau^{-\alpha}$, with $\alpha=1.1$, consistent with the range found by \cite{Padovani1993}, who calculate $0.9\lesssim \alpha \lesssim 1.3$ for elliptical galaxies. Observable jet properties, if dictated by entrainment, are thus strongly affected by stellar population age, as discussed below. 

\addedit{As we focus here on stellar mass-loading, we have neglected the ISM as another source of entrainment. Inclusion of an ISM mass-loading term should increase the fraction of jet power available for direct feedback in the galaxy, as opposed to preventative feedback on larger scales. However, the significance of the ISM contribution may decrease with time, depending on how easily the jet cocoon clears the ISM.} 

\subsection{Global energetics}\label{sec:disc_energetics}
As jets are thought to be \removeedit{the}\addedit{fundamental to kinetic}-mode feedback in galaxies, understanding how their energy and momentum are made available to the ISM is an essential question. Within the framework of our model, we find that it is relatively easy to decelerate \addedit{most AGN} jets, making their kinetic power available for \addedit{direct feedback (mediated by the coupling efficiency to the surrounding gas)}\removeedit{coupling to the host galaxy}.

\cite{Perucho2014} found that stellar mass-loading \addedit{from old stellar populations} could decelerate jets up to $P\sim10^{42-43}~\rm erg~s^{-1}$, \addedit{while we find $\bar{P}_{\rm decel}=10^{43.1}~\rm erg~s^{-1}$ for our most massive galaxy sample at $\log \tau=10$} (the top panel of Fig.~\ref{fig:power_plots} multiplied by $M_*$)\addedit{, closely matching the upper end of their findings.} \addedit{\cite{Perucho2014} used a Nuker stellar brightness profile normalized such that the total mass-loss out to 20~kpc is ${\sim}0.4 ~\Msun~\rm yr^{-1}$, within a factor of ${\sim}$ few from ours depending on the galaxy mass. However, their jets only propagate for ${\sim}2~$kpc in the simulation box.}

\removeedit{although their adopted mass-loading rates are ${\sim}$1\,dex lower than the ones found here. Thus, our finding that jets up to $P\sim10^{44}~\rm\,erg\,s^{-1}$ (the top panel of Fig.~5 multiplied by $M_*$) are significantly decelerated by stellar mass-loading in massive galaxies ($M_*=10^{11.7}\,\Msun$) is consistent with their results, accounting for the difference in MLR.}

\removeedit{Although higher mass galaxies are more likely to host a jet that escapes the galaxy undecelerated (Fig.~4),} \addedit{High mass} galaxy populations receive more feedback power from decelerated jets, even normalized by stellar mass (Fig.~\ref{fig:power_plots} top panel), with a $\sim$1~dex increase over $\sim$1~dex in stellar mass. Thus we find that jet feedback should be more impactful as stellar mass increases, agreeing with a long body of work focusing on radio mode AGN feedback in massive galaxies.

To constrain the actual jet power available for feedback, the bottom panels of Fig.~\ref{fig:power_plots} shows the global fraction of jet power $f_P$ available for feedback\removeedit{ (i.e. fraction of jet power measured to be ``decelerated'', as determined by our threshold)}, across a range of stellar masses and as a function of stellar age. The fraction decreases for older stellar populations, as the highest-power jets progressively escape more easily. $f_P$ is part of a correction factor for global energetics calculations \citep[e.g., those of][]{Heckman2023, Kondapally2023, Heckman2024, Igo2025, Kondapally2025}. \addedit{$f_P$ distinguishes between the power available for direct feedback in the galaxy, as opposed to that available for preventative feedback throughout the halo.}

For our fiducial model of $\theta=3\degree,~\Gamma_j=3$, we find $\gtrsim25\%$ of jet kinetic power available for feedback even at old stellar population ages, and increases close to 100\% for young stellar populations ($\tau\leq10^8~\rm yrs$). In practice, to be applied quantitatively to a global energetics calculation \addedit{for direct feedback}, $f_P$ must be integrated over jet opening angle and Lorentz factor distributions, along with stellar population ages over the local galaxy population. Figure~\ref{fig:theta_variation} shows the variation in $f_P$ with jet opening angle, which increases with $\theta$ as expected. Between opening angles of 1-5$\degree$, we find an increase in $f_P$ by approximately a factor of 2. Smaller opening angles decrease $f_P$ at younger stellar population ages, limiting the availability of kinetic power for feedback inside galaxies.\removeedit{ As distributions of $\theta, ~\Gamma_j$, and stellar population ages are poorly constrained, we do not perform them here. }

Interestingly, this correction factor is essentially independent of stellar mass, and thus also free from variations in the galaxy stellar mass function. \addedit{Therefore, while integration over $\theta, \Gamma_j$ might be necessary for a calculation of the power available for direct feedback, the shape of the stellar mass function is less important.} \removeedit{Still, preventative feedback (see Figures 5-7 of Igo et al., 2025) should remain efficient in galaxies of $M_*\lesssim10^{11.2}\,\Msun$.} We thus suggest that stellar mass-loading is an efficient process of enabling \addedit{direct} jet-mode feedback, particularly \removeedit{in lower-mass galaxies and }in galaxies with younger stellar populations (which could be more significant at higher redshifts). \

\subsection{Jet morphology}\label{sec:disc_morph}
Initially, the Fanaroff-Riley \addedit{morphological} classification was thought to correspond to jet power, with FR-II jets having higher radio luminosities \citep{Ledlow1996, Ghisellini2001}. More recent studies with homogeneous samples have complicated our picture of the FR-I/II dichotomy \citep[see e.g.,][and references within]{Best2009,Mingo2019,Magliocchetti2022, Mingo2022}. FR-IIs are now found in both low and high excitation radio galaxies \citep[with varying frequencies in different surveys, e.g.,][]{Buttiglione2010, Best2012, Janssen2012, Capetti2017, Mingo2022}. How jet launching mechanisms, host galaxy properties, and environments, combine to form FR-II jets is still poorly understood. 

\begin{figure}
    \centering
    \includegraphics[width=\linewidth]{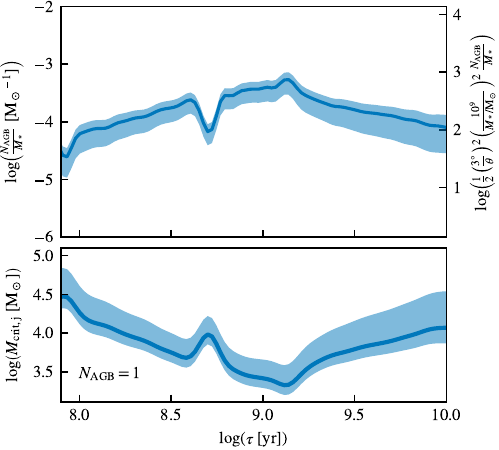}
    \caption{Top panel: Number of AGB stars, $N_\mathrm{AGB}$, as a function of stellar population age, scaled by stellar mass. The right-hand axis shows the expected number of AGB stars in a bidirectional jet with a $3\degree$ opening angle and stellar mass of $10^9\,\Msun$. The bottom panel shows the critical stellar mass necessary to expect 1~AGB star within a jet of opening angle 3$\degree$, enclosed in the jet.}
    \label{fig:agb_stochasticity}
\end{figure}

\addedit{As FR-I/II jets differ in terms of deceleration, with FR-I jets being decelerated on kpc-scales,} we can ask whether our model of mass-loading alone can explain the FR-I/II dichotomy, in the sense that \addedit{FR-I jets are decelerated within the galaxy, while FR-II are not}\removeedit{FR-II jets successfully carry their energy out of the galaxy and FR-I jets fail to do so} (i.e., can we explain the dichotomy as a consequence purely of mass loading for radio galaxies drawn from the single parent population described in eq.~(\ref{eq:edd_frac}), located in galaxies of different stellar mass). 

To test this hypothesis, we thus label FR-II jet candidates as those that exceed our critical power threshold. Following that line of reasoning, $1 - f_{\rm decel}$, as shown in Figure~\ref{fig:f_stopped}, is the expected fraction of FR-II jets within the \addedit{local}\removeedit{broader} jet population. \addedit{The top panel of Fig.~\ref{fig:f_stopped} shows the expected escape fraction considering only the $\lambda$ range of \cite{Igo2024}, with horizontal dashed lines marking the FR-II fraction in the \cite{Igo2024} sample. \cite{Igo2024} found 10-30\% of the radio AGN to have FR-II-like morphologies, however they do not identify clear trends with host stellar mass.} More statistics and higher resolution radio surveys would be needed to\removeedit{analyze any trends with host galaxy stellar mass, i.e. to} measure the escape fraction explicitly as a function of stellar mass (with the escape fraction assumed to be equivalent to $1 - f_{\rm decel}$.) \addedit{Our predictions are consistent with their findings for old stellar populations, allowing for a reasonable range of $\theta$ (as well as $\Gamma_j$, which is not shown.)}

\addedit{Our model predicts that jets escape more easily from old galaxies. Recent studies have found a large number of FR-IIs hosted in massive red LERGs. For instance, \cite{Capetti2017} states that 90\% of FR-IIs in their sample were hosted in LERGs, which were indistinguishable from the hosts of FR-Is by color and concentration. 77\% of the FR-II sample of \cite{Mingo2022} were also hosted in LERGs. Hence, our finding that older galaxies more easily produce FR-IIs is in line with the LERG subsample of FR-IIs.}

\addedit{Our model does not easily explain the population of FR-II HERGs. However, as such AGN are likely hosted in more disk-like galaxies \citep[e.g.,][]{Miraghaei2017}, our assumption of a spherical galaxy may not be accurate. The angle between the jet and the disk (i.e., whether the jet points into or perpendicular to the disk) will strongly affect the amount of stellar mass-loss the jet is exposed to, which is a parameter we do not account for. We leave modeling of FR-II HERGs (i.e., a younger stellar population and bulge) to future work.}

\addedit{Looking at the extrapolated sample (bottom panel of Fig.~\ref{fig:f_stopped}), which we expect to include sources not identified by the LoTTS survey,} we find that mass-loss efficiently decelerates the vast majority ($\sim99\%$) of jets in all cases, and particularly so for lower-mass galaxies, meaning that the \addedit{extrapolated ``true''} fraction of FR-II sources should be small, ${\sim}1\%$ for high mass galaxies. Based on this mass-loading model, we would further expect to observe FR-II jets predominantly in more massive galaxies with older stellar populations. \addedit{However, the mass-dependence here is very sensitive to the mass- and flux-complete measurements of AGN incidence.}

\removeedit{Our prediction is in contrast with the traditional observational picture of FR-II jets, which are hosted in less massive galaxies with higher star formation rates, indicating a younger stellar population. Recent studies have continued to uphold these findings (see e.g., Wright2010, Miraghaei2017, Mingo2019). However, FR-II jets have also been found in massive, red galaxies (Capetti2017).}

\removeedit{Performing a quantitative comparison is difficult due to the varying samples used in observations and the selection effects of each. Broadly speaking, however, our toy model does not replicate the observed trends in FR-II host galaxies. Our simple hypothesis is thus insufficient to explain the difference between FR-I and FR-II sources. There are several possible explanations: firstly,} \addedit{Our model captures several important features of the FR-II population, but we also neglect several contributing factors:}

\begin{itemize}
    \item FR-I/II sources may have jet launching variations, e.g., in opening angle, Lorentz factor, and accretion rate distribution \citep[see e.g.,][]{Best2005, Baldi2018, Whittam2018, Smith2025}.
    \item \addedit{As mentioned above}, host galaxy morphologies may be important for FR-IIs. We have assumed a quasi-spherical stellar population, \addedit{but FR-II HERGs are likely hosted in diskier galaxies. In such cases, the geometric assumptions in Sect.~\ref{sec:meth_jet} will fail.}\removeedit{Since FR-II are generally observed in HERGs, which are typically diskier (Miraghaei2017), the assumed geometry may fail. In such cases} \addedit{Moreover}, the galaxy bulge mass and stellar mass are no longer equivalent, such that our relation in eq.~\eqref{eq:lambda_crit} might not hold.
    \item \addedit{If unaccounted for factors (e.g., magnetic fields) maintain collimation and a parabolic jet profile on galactic-scales \citep[for instance, Cygnus A is parabolic out to 24~kpc,][]{Nakahara2019}, the covering fraction of the jet across the galaxy may be altered depending on the measured conical opening angle. Collimation will also prevent flaring of the jet, which we also neglect.}
    \item We neglect \removeedit{other contributing factors of jet deceleration, such as} turbulent mixing at the jet boundary with the ISM.
    \item Our single threshold for deceleration (a reduction in $\Gamma$ by a factor of 2) may also be insufficient, with high Lorentz factor jets meeting this threshold while remaining significantly relativistic and carrying the bulk of their energy out of the galaxy \citep[blazars are thought to potentially have $\Gamma_j$ up to 40, e.g.,][]{Saikia2016}. Sect.~\ref{sec:disc_thresh} and Appendix~\ref{sec:app_thresh} discuss our deceleration threshold in greater detail.
\end{itemize} 

\begin{figure}
    \centering
    \includegraphics[width=\linewidth]{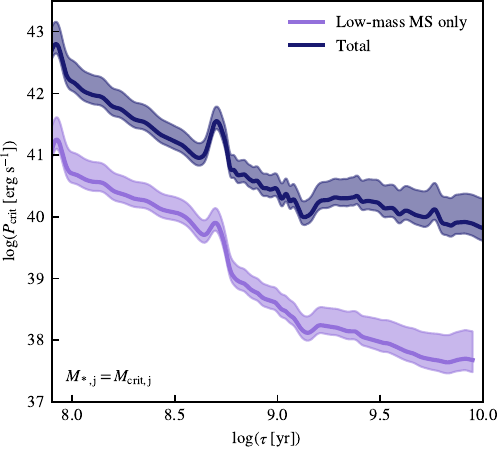}
    \caption{Critical jet power stopped by stellar populations of mass $M_{\rm crit}$ from the bottom panel of Figure~\ref{fig:agb_stochasticity}. The light purple line shows the critical jet power if only mass loss from low-mass MS stellar populations is considered. The dark blue line shows the addition of one AGB star (i.e. using the total MLR with $M_*=M_{\rm crit}$ \addedit{from the bottom panel of Fig.~\ref{fig:agb_stochasticity}}.)}
    \label{fig:p_crit_1_agbstar}
\end{figure}
 
\subsection{Stochasticity}\label{sec:disc_stoch}
Stellar mass-loss in galaxies is dominated by AGB stars, which are relatively rare compared to main sequence stars. Our model implicitly assumes a secular mass-loading process that fills the volume of the jet. If mass-loading from AGB stars becomes a stochastic process, then the validity of the model is in question.

If mass-loading becomes stochastic, the physical distribution of mass-loss sources may also become important, especially in low-mass galaxies. \cite{Komissarov1994} argued that a large number of mass-loss sites are necessary for the stellar tails to overlap and decelerate the jet uniformly. In low-mass galaxies with few AGB stars contributing to the MLR, the positioning of stellar sources and dynamics of mass-loss become more important.

To assess the question of stochasticity, we calculate the number of AGB stars per stellar mass at any given time, found by integrating the IMF over the appropriate mass bounds (i.e. what range of stellar mass stars are in the AGB phase). The top panel of Figure~\ref{fig:agb_stochasticity} shows the result of this integral normalized by stellar mass. We also scale this relation by galaxy stellar mass and jet opening angle, as shown on the right axis. For higher mass galaxies, we expect ${\gg}1$~AGB stars within the path of the jet (e.g., ${\sim}10^4$ AGB stars for a $M_*=10^{11} \,\Msun$ galaxy). \addedit{However, }in low-mass galaxies, e.g., a $M_*=10^9\,\Msun$ galaxy as labeled on the right hand axis, a secular treatment of AGB mass loss may not be warranted. If only ${\lesssim}100$ AGB stars reside in the path of the jet, the Poisson uncertainty in that number begins to approach the number itself and a secular treatment is not appropriate.

Although AGB stars have the largest mass loss rates, the galaxy stellar mass is dominated by the main sequence stars. \addedit{The lower panel of Fig.~\ref{fig:agb_stochasticity} shows the stellar mass (almost entirely in MS stars) enclosed in a jet to host 1~AGB star.} Especially in the context of AGB stochasticity, we are thus interested whether there are situations in which a jet might be decelerated by these low-mass MS stars before reaching even a single AGB star. The light purple curve of Figure~\ref{fig:p_crit_1_agbstar} shows the critical jet power that the main sequence stars can decelerate before reaching the first AGB star. That curve thus shows the maximum jet power that main sequence stars decelerate before AGB stars take over, which is independent of actual stellar mass or opening angle, \addedit{assuming that AGB stars are distributed similarly to the MS stellar population.} The dark blue line shows $P_{\rm crit}$ with the addition of a single AGB star, leading to a 2-3~dex increase in $P_{\rm crit}$.

Jets that would be slowed before reaching an AGB star are very weak in older stellar populations (${\lesssim} 10^{38}~\rm erg\,s^{-1}$). We thus find that, in galaxies with old stellar populations, evolved stars always play the dominant role in stellar mass-loading, despite their small numbers. However, galaxies with younger stellar populations could also host radio AGN, especially at higher redshift. In galaxies with 100~million year old stellar populations, a jet with power $\lesssim10^{41}\,\rm erg\,s^{-1}$ could be decelerated before reaching even a single AGB star due to the combination of AGB star rarity and the relative strength of main sequence winds at that age. A jet of that power with $\lambda = 10^{-3}$ would be hosted in a galaxy of $M_*\sim4\times10^8\,\Msun$. 

Although such galaxies seem very low mass to host a radio AGN, small samples of such sources have been observed \citep[][]{Greene2006,Reines2011, Sartori2015,Reines2020}. Detecting such sources is also difficult due to the non-linear scaling between kinetic power and radio luminosity, as observational selection effects may mask a large population of such sources. Moreover, as SMBH cycle through accretion modes and associated Eddington fractions, many galaxies may host weak jets at very low $\lambda$. \addedit{However, at} these masses stellar feedback is thought to be highly efficient on its own (at least in galaxies of young stellar population ages), indicating that jet feedback is not necessary. 

Even a single AGB star dominates over a \addedit{MS} stellar population of ${\sim}10^5\,\Msun$, showing that AGB mass-loading is dominant over all stellar population ages. However, it may fail when the host stellar mass is small enough that there are few AGB stars ($\lesssim10^8\,\Msun$), which may occur in dwarf galaxies. If such galaxies do host radio AGN (see above), the stochasticity of AGB stars might introduce a significant random variability in jet deceleration efficacy, because of the outsized influence of individual AGB stars. Additionally, a consideration of the time variability in AGB wind MLRs would become important in such cases. AGB winds become significantly stronger in the late stages of the AGB branch \citep[i.e. AGB ``superwinds'' with $\dot{M} \gtrsim 10^{-5}\rm~\Msun~yr^{-1}$ lasting mere hundreds to thousands of years, see e.g.,][Sect.~3.1.2, and references within. Here, we have averaged over such superwinds.]{Hofner2018} Observational signatures of AGB mass loss could help distinguish AGB vs \removeedit{mass}\addedit{main} sequence mass loading \citep[e.g., knots in jets, some of which could be mass loading sites, as argued by][for the case of Cen A]{Hardcastle2003, Wykes2015, Wykes2019}

\subsection{Radio AGN incidence}\label{sec:disc_cav_incidence}
We study only local radio sources, but radio AGN incidence is known to evolve with red-shift; \cite{Delvecchio2022} found that for a fixed stellar mass, the jet kinetic luminosity density increases out to $z=2$. \cite{Zhu2023} also observed a varying dependence of radio AGN incidence on stellar mass with redshift, with a shallower $M_*$-incidence relation at higher redshift. The properties of radio loud AGN also change; while locally, few radio AGN also have strong emission lines, at higher redshift AGN can be both radio loud and have strong emission lines \citep{Kauffmann2008}. Moreover, local LERGs and HERGs are differentiated not only by accretion mode but also galaxy properties \citep[in stellar mass, SFR, \addedit{and color distributions}; see][]{Chen2013, Magliocchetti2022}, but this trend becomes less clear at $z>1$.

This redshift evolution of the radio AGN-hosting galaxy population is particularly interesting as we see a major age evolution in the galaxy-wide stellar MLR, and thus $P_{\rm crit}$. Although radio jets in the local Universe typically pass through older stellar populations (i.e. ages greater than a ${\sim}$few~Gyr), more distant radio AGN are more likely to be hosted in galaxies with younger stellar populations, capable of slowing down more powerful jets with entrainment. Stellar mass-loading should thus remain equally, if not more, important at higher redshifts. \addedit{It is worth noting that at higher redshifts, as radio AGN are hosted in younger galaxies, ISM contributions to mass-loading might be increasingly important as well.}

Moreover, observed radio AGN incidence is highly affected by selection biases (e.g., in surface brightness). Future observations may expand our picture of radio AGN hosts. The increasing number of detections of radio AGN signatures in low-mass galaxies raises the possibility that radio jets may be hosted in dwarf galaxies where main sequence stellar winds are more important (see Sect. \ref{sec:disc_stoch}). On the other hand, there may be a population of undetected, extremely low $\lambda$, jets. There is even the possibility that the Milky Way's own Sgr A* hosts an extremely weak radio jet \citep[][]{EHT2024, Chavez2024}, with the Fermi bubbles also suggesting that it may have hosted a more powerful jet in the past \citep[e.g.,][]{Su2010, Guo2012}. \addedit{The effect of mass-loading on apparently quiescent SMBHs hosting weak radio jets in disk galaxies might depend on the angle between the jet and the disk \citep[i.e. whether the jet points into the plane of the disk or not, see e.g.,][]{Girdhar2022} and would require additional model parameters.}

\removeedit{Late-type galaxies with apparently quiescent SMBHs hosting weak radio jets would be particularly affected by mass-loading in our model. In such cases, the jet angle relative to the disk might be significant due to the non-spherical distribution of stars in the disk, depending on the jet length, see e.g., A. Girdhar et al. 2022, but we would expect such jets to be easily frustrated in any case.}

\subsection{Caveats}\label{sec:disc_thresh}
Our model of jet deceleration (Sect. \ref{sec:meth_jet}) is based on conservation of momentum. It provides first-order insights into jet propagation with mass-loading included, while neglecting the detailed physics. In the derivation of our ``deceleration threshold'', eqs. \eqref{eq:P_crit} and \eqref{eq:lambda_crit}, we make a variety of assumptions: the ultra-relativistic limit, a $M_{\rm BH}-M_*$ relation, a mass-loading efficiency $\epsilon=1$ (i.e. the entirety of the stellar wind is entrained). Additionally, we do not consider the spatial distribution of mass-loading, implicitly assuming a spherically symmetric system. A stellar density profile would be required to calculate quantities such as a deceleration length. \addedit{Sect.~\ref{sec:disc_morph} discusses additional limitations of our study.}

\removeedit{In order to more accurately replicate observed morphological differences between FR-I and FR-II jets, a different approach to the threshold may be necessary. As is, the requirement for the Lorentz factor to be halved may allow jets to be counted as ``decelerated'' without reaching speeds where the jet and ISM would actually interact. A jet with initial $\Gamma_j=10$ could meet our deceleration threshold by reaching $\Gamma=5$, while remaining decoupled from the galaxy. Appendix Sect.6.2 compares these methods for our fiducial case, however, it is possible that the difference between FR-I/II sources is a combination of power, opening angle, and large $\Gamma_j$, as we find that the jet power distribution as a function of galaxy mass alone does not produce a prediction consistent with the observed distributions of FR-II jets, regardless of the method adopted.
}
\section{Conclusion}\label{sec:conclusion}
We have presented a straightforward model of jet deceleration by entrainment of stellar winds (Sect.~\ref{sec:meth_jet}), along with calculations of stellar mass loss rates (Sect.~\ref{sec:meth_ml}). In particular, we present novel calculations of main sequence thermal MLRs derived from X-ray luminosities (Fig.~\ref{fig:bootstrap_Lx} and Sect.~\ref{sec:ml_ms}.) 

Paired with radio AGN incidence data from \cite{Igo2025}, we calculate the amount of kinetic feedback made available by mass-loading-driven deceleration. We find the following:

\begin{itemize}
    \item{Galaxy-wide stellar MLRs are dominated by AGB stars, and evolve as $\tau^{-1.1}$ with stellar population age. Main sequence winds contribute 5-10\% of the total MLR for stellar populations aged $\tau\sim10^{8.5-9}\,{\rm yrs}$.}
    \item{Stellar mass-loading\removeedit{is} \addedit{can be} highly efficient, such that when integrated over radio AGN incidence, at least $\gtrsim25\%$ of global jet power is available for \addedit{direct} feedback within \addedit{early-type} host galaxies \addedit{at low redshifts (Fig.~\ref{fig:power_plots})}. \addedit{The coupling efficiency between a decelerated jet and the surrounding ISM will further reduce the thermal energy injected into the ISM.}}
    \item \addedit{We find that $f_P$, the fraction of jet power available for direct feedback, is independent of stellar mass, and largely invariant to the shape of the radio AGN incidence function at low jet $\lambda$ (i.e., extrapolation of the incidence function from \cite{Igo2025} does not strongly impact $f_P$.)}
    \item \addedit{Our model is able to match the FR-II fraction in \cite{Igo2024}, with a reasonable range of jet parameters. Our stellar mass-loading model cannot explain the population of FR-II HERGs, however, this may be because our model does not accurately apply to non-elliptical galaxies. We predict that if all massive galaxies host low level radio AGN activity \citep[e.g.,][]{Sabater2019}, the ``true'' fraction of jets that escape their host galaxies undecelerated should be significantly lower, ${\sim}1\%$.}
    \removeedit{We find that mass-loading cannot, on its own, explain the existence of FR-II radio galaxies in lower mass or late-type galaxies, if the underlying jet power distribution is the same for both radio galaxy populations. Other factors, such as intrinsic variations in opening angle or speed, or galaxy morphology and age, likely play a role.}
    \item{In low-mass dwarf galaxies, where AGB mass-loading may become highly stochastic, MS stars in young stellar populations ($\tau\sim10^8$~yrs) may provide an effective mechanism to slow low power jets. As a corollary, we find that low power jets may be slowed down effectively by MS winds even before reaching the first AGB star inside the jet, especially for younger stellar populations, where the critical power for this to occur approaches $P_{\rm crit} \lesssim 10^{41}\,{\rm ergs/s}$, independent of galaxy and black hole mass and jet opening angle.}
\end{itemize}

In reality, galaxies host stellar populations of varying ages. Our results can thus be integrated over an arbitrary stellar age distribution to form a more accurate MLR. We discuss additional caveats to the model in Sect.~\ref{sec:disc_thresh}. Ultimately, we find that stellar wind mass loading provides an efficient means of decelerating jets across most of the distribution of jet powers observed in galaxies. Furthermore, when calculating \addedit{the contribution of jets to direct} kinetic feedback, corrections to the integrated jet power released by accreting black holes are \removeedit{below}\addedit{within} 1~dex for typical jet parameters. Future studies could account for more detailed jet and galaxy models in order to create a more accurate correction factor for global energetics calculations, as well as producing further refinement of the model.

\begin{acknowledgements}
\addedit{The authors thank Eric Hooper and Marsha Wolf for useful discussions, and the anonymous referee for a thoughtful report that improved the work. TMO acknowledges that this material is based upon work supported by the National Science Foundation Graduate Research Fellowship Program under Grant No. 2137424. Any opinions, findings, and conclusions or recommendations expressed in this material are those of the author(s) and do not necessarily reflect the views of the National Science Foundation. Support was also provided by the Graduate School and the Office of the Vice Chancellor for Research at the University of Wisconsin-Madison with funding from the Wisconsin Alumni Research Foundation. SH and TMO acknowledge support from NASA grant 21-ATP21-0048.} ZI acknowledges the support by the Excellence Cluster ORIGINS which is funded by the Deutsche Forschungsgemeinschaft (DFG, German Research Foundation) under Germany´s Excellence Strategy – EXC-2094 – 390783311. 
\end{acknowledgements}

\bibliography{sample631}{}
\bibliographystyle{aasjournalv7}
\appendix
\section{Deceleration criteria}
\label{sec:appendixA}
\subsection{Derivation}\label{sec:app_deriv}
Here, we present a slightly more formal (though still in the ultra-relativistic limit) derivation of the critical mass loading rate to decelerate a jet of power $P_{j}$ and initial Lorentz factor $\Gamma_{j}$, taking into account the increase in inertia of the swept mass in the jet frame due to the work done on the swept up mass.

In the frame of the jet, any instantaneously swept up mass will be accelerated to the same Lorentz factor $\Gamma$ as the jet material itself. Particles will, on average, be endowed with a random Lorentz factor of $\langle \gamma \rangle = \Gamma$, increasing their effective inertia by the same factor. Thus, in the observer's frame, the total momentum flux down the jet will be given by the jet material itself and the integral over the swept up mass:
\begin{equation}
    \dot{\Pi}_{\rm tot}=\dot{\Pi}_{j} + \Gamma(\dot{M_j})  c\int_{0}^{\dot{M_j}}d\dot{M}_j'~ \Gamma(\dot{M}'_j).
    \label{eq:mom_con}
\end{equation}
where material will be swept up along the jet at decreasing $\Gamma$. Here, we consider the mass loading rate $\dot{M}_j$ within the jet as a Lagrangian variable along the jet. Hence, $\Gamma(\dot{M}_j=0)=\Gamma_{j}$ at the base of the jet and $\Gamma$ decreases with increasing $\dot{M}_j$ measured along the jet.

If we neglect any work done on the jet plasma itself, we can write
\begin{equation}
    \dot{\Pi}_{j}=\frac{P_{j}}{c}\frac{\Gamma}{\Gamma_{j}}.
\end{equation}
Taking two derivatives of eq.~(\ref{eq:mom_con}) with respect to $\Gamma$, we can write the equation of motion as
\begin{equation}
    3\frac{d\dot{M}_j}{d\Gamma} + \Gamma \frac{d^{2}\dot{M}_j}{d\Gamma^2} = 0,
\end{equation}
the solution of which is elementary: 
\begin{equation}
    \dot{M}_j=\frac{\dot{M}_{1}}{\Gamma^2} + \dot{M}_{0},
\end{equation}
with two constants of integration $\dot{M}_{0}$ and $\dot{M}_{1}$, which can be solved for using the following initial conditions:
\begin{equation}
    \left.\dot{M}_j\right|_{\Gamma=\Gamma_{j}}=0,
\end{equation}
and
\begin{equation}
    \left.\frac{d\dot{\Pi}}{d\Gamma}\right|_{\Gamma=\Gamma_{j}}=\frac{P_{j}}{c\Gamma_{j}} + \Gamma_{j}^{2}c\frac{d\dot{M}_j}{d\Gamma}=\frac{P_{j}}{c\Gamma_{j}} - 2c\frac{\dot{M}_{1}}{\Gamma_{j}}=0.
\end{equation}
Incorporating those conditions, we find
\begin{equation}
    \dot{M}_j=\frac{P_{j}}{2\Gamma_{j}^{2}c^{2}}\left(\frac{\Gamma_{j}^{2}}{\Gamma^{2}} - 1 \right),
\end{equation}
and, conversely,
\begin{equation}
    \Gamma=\frac{\Gamma_{j}}{\sqrt{1 + \frac{2\Gamma_{j}^{2}c^{2}\dot{M}_j}{P_{j}}}}.
\end{equation}

We can solve for the total mass loading rate for which half of the total momentum flux is transferred to the swept-up mass, which occurs when
\begin{equation}
    \Gamma c\int_{0}^{\dot{M}_{j,\rm crit}}d\dot{M}_j'\Gamma(\dot{M}_j')=\frac{P_{j}}{2\Gamma_{j}^2 c}.
\end{equation}
From there we find
\begin{equation}
    \dot{M}_{j,\rm crit} \equiv \frac{3}{2}\frac{P_{j}}{c^{2}\Gamma_{j}^{2}} \ \ \ \ \ \ {\rm and} \ \ \ \ \ \ P_{\rm crit}\equiv\frac{2c^2 \Gamma_{j}^{2}\dot{M}_j}{3}.
\end{equation}

If we transform to the MLR of the galaxy $\dot{M}$ using eq.~\eqref{eq:meth_frac}, $P_{\rm crit}$ then takes the form
\begin{equation}
    P_{\rm crit}=\frac{\Gamma_{j}^2 \theta^2 c^2\dot{M}}{3}.
    \label{eq:P_crit_dynamical}
\end{equation}

\subsection{Deceleration thresholds}\label{sec:app_thresh}
Throughout the paper, we adopt a simple approximation for what we define as the deceleration threshold for mass loading. Here, we explore variations in that definition and their effect on our findings:

\begin{enumerate}
    \item \textbf{Method (\addedit{1})}: Dynamical $\dot{\Pi}_* = \frac{1}{2}\dot{\Pi}_{\rm tot}$, considering the change in $\langle \gamma \rangle$ as the jet propagates, as presented in \addedit{eq.~\ref{eq:P_crit}, and derived more fully in }Appendix~\ref{sec:app_deriv}. \removeedit{presents a more rigorous version, . We still require a 50\% momentum transfer. Eq.~\eqref{eq:P_crit_dynamical} varies by from Method (1) by a small factor of order unity (1/2 to 2/3).}
    \item \textbf{Method (\addedit{2})}: $\Gamma = 1$. We may define \textit{complete deceleration} as $\Gamma=1$, such that

    \begin{equation}
        P_{\rm crit} = \frac{\dot{M} \theta^2 c^2 \Gamma_j^2}{\Gamma_j^2 - 1}.
    \end{equation}

    \item \textbf{Method (3)}: $\theta \lesssim \frac{1}{\Gamma}$. If the opening angle of the jet exceeds the causal angle of $\theta_{\rm causal} \approx 1/\Gamma$, the jet will propagate ballistically along the conical geometry until it is decelerated to $\Gamma \lesssim 1/\theta$. Beyond this point, the jet can come into transverse pressure balance by expanding laterally, leading $\theta$ to increase, which leads to rapid deceleration \addedit{\citep[e.g.,][]{heinz:99, Komissarov2012, Smith2025}}. However, this requirement is generally already satisfied for typical AGN jets, i.e. $\Gamma_j\sim$ few and $\theta \sim \rm few \degree$.    
\end{enumerate}

\begin{figure}
    \centering
    \includegraphics[width=0.65\linewidth]{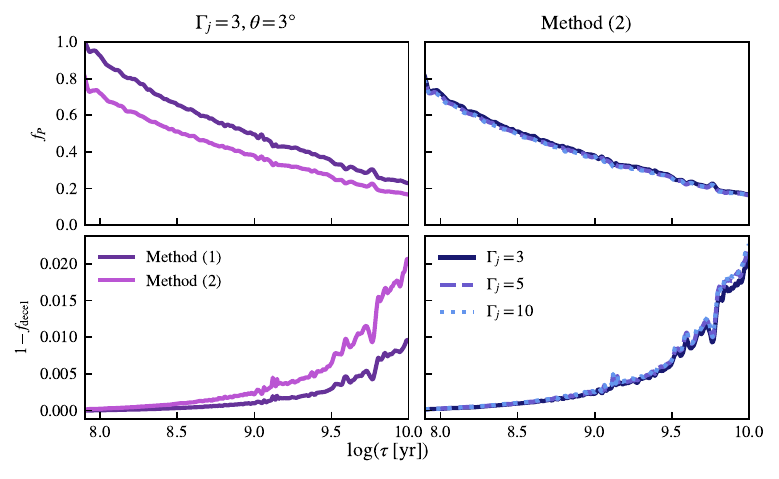}
    \caption{Comparison of the deceleration thresholds for \textit{top:} fraction of kinetic power (i.e., Fig.~\ref{fig:power_plots}) in decelerated jets, and \textit{bottom:} fraction of jets (i.e., Fig.~\ref{fig:f_stopped}) that escape their hosts. On the left, we compare Methods (1) and (2) as listed in Sect.~\ref{sec:app_thresh}, showing results for $\theta=3\degree,~\Gamma_j=3,~\rm and~M_*=10^{11.7}~\Msun.$ On the right, we vary the initial $\Gamma_j$ for Method (\addedit{2}) only, still with $\theta=3\degree\rm and~M_*=10^{11.7}~\Msun$. In both plots, we use the extrapolated incidence function from \citet{Igo2025}.}
    \label{fig:thresholdcomparison}
\end{figure}

The left side of Figure~\ref{fig:thresholdcomparison} shows important figures from the main text re-calculated for the thresholds listed above. The choice of deceleration threshold provides a ${\lesssim}10\%$ variation in $f_P$ among old stellar populations, well within the bounds of our uncertainties in Fig.~\ref{fig:power_plots}. Stringently requiring that $\Gamma = 1$ for deceleration (pink line in Fig.~\ref{fig:thresholdcomparison}) provides an upper limit on the percent of jets we expect to escape their hosts, but still remains below ${\sim}2\%$ among the highest mass galaxies in our sample ($M_* = 10^{11.7}~\Msun$). As shown in the right panels of Fig.~\ref{fig:thresholdcomparison}, the deceleration fraction is essentially independent of $\Gamma_j$ for Method (3).

\section{Low-mass main sequence stellar winds}
\label{sec:appendixB}
\addedit{Here, we discuss our model for low-mass MS winds. We examine briefly the assumption that $\dot{M}\propto L_X$ in Sect.~\ref{sec:appb_lx_mdot} and provide further references for interested readers. In Sect.~\ref{sec:appb_dataprep} we provide more details about how we prepare the data from \cite{Nunez2016, Booth2017, SanzForcada2010} as shown in Fig.~\ref{fig:bootstrap_Lx}.}

\subsection{Connection between $L_X$ and $\dot{M}$}
\label{sec:appb_lx_mdot}
\addedit{We cannot directly measure MLRs from low-mass MS thermal winds. H~I Lyman~$\alpha$ absorption detections have only produced ${\sim}20$ indirect measurements \citep[][]{wood_new_2021}, which are insufficient for a statistical relation between age and MLR. Certain other methods \citep[such as transient absorption features,][]{Jardine2019} can only be used for fast-rotating stars. Because of these challenges, we seek a more easily observed proxy for mass loss rate, motivating our use of the X-ray luminosity. $L_X$ and $\dot{M}$ are presumed to be related through stellar activity }\citep[for a discussion of $L_X$ and its association with magnetic flux and stellar age, see][]{Pevtsov2003, Testa2015}. Nevertheless, how exactly the two quantities relate remains debated. \cite{cohen_independency_2011} find little dependence of the solar MLR on the Sun's X-ray flux (i.e. the MLR does not vary strongly between solar minimum and maximum), arguing that the MLR is determined by the open magnetic flux of the Sun, rather than the closed magnetic flux that the X-ray luminosity measures. However, other sources studying a larger range of stars do observe a $\dot{M}-L_X$ relationship. Studies of solar-like stars by \cite{wood_new_2005} found $\dot{M} \propto F_X ^ {1.34}$, for $F_X < 3.9 \times 10^{27} \mathrm{erg ~s^{-1}} ~ (8 \times 10^5 \mathrm{erg~cm^{-2} ~s^{-1}}$). Yet, later work by \cite{wood_new_2021} find a loose relationship of $\dot{M} \propto F_X ^ {0.77}$ in their study of M-dwarfs. 

Whether and how rotation speed ($\Omega$) contributes to mass loss rate is debated also. For example, \cite{johnstone_stellar_2015} found mass loss rates to be proportional to $\Omega ^ {1.33}$, whereas \cite{cohen_independency_2011} do not find a correlation throughout the solar cycle. We implicitly incorporate the well-documented rotation-age dependence \citep[see e.g.,][]{Mamajek2008, Pass2022, Engle2023} through our age-$L_X$ relation. 

\subsection{$L_X$ data preparation}
\label{sec:appb_dataprep}
\addedit{Each X-ray luminosity dataset that we used was measured in varying X-ray bands, so we convert each to} 0.1-3~keV with an APEC model in WebPIMMS\footnote{WebPIMMS can be accessed here: \url{https://heasarc.gsfc.nasa.gov/cgi-bin/Tools/w3pimms/w3pimms.pl}}, consistent with ROSAT measurements of the solar X-ray luminosity. In particular, we examine $L_X$ normalized by stellar surface area (i.e. the X-ray flux through the stellar surface), because that quantity is consistent across spectral types ranging from M to F \citep[][]{Schmitt2004}. The details of each conversion are below:

\begin{itemize}
    \item \textbf{\cite{Nunez2016}}: \addedit{X-ray data originally measured in 0.5-7.0~keV.} We calculate stellar radius from the given mass with $R = M^{0.8}$ in solar units, which is broadly consistent with many mass-radius relations in the literature \addedit{\citep[see discussion in][and references within]{Eker2024}.} \removeedit{ and relations from e.g., (Patterson1984,Gimenez1985,Demircan1991) who find fits to $R = \alpha M^\beta$ with $\beta$ ranging from 0.74 to 0.945, and $\alpha$ from 0.89 to 2}
    \item \textbf{\cite{Booth2017}}: \addedit{X-ray data originally measured in 0.2-2.0~keV. Data was already available in terms of surface flux.}
    \item \textbf{\cite{SanzForcada2010, SanzForcada2011}}: \addedit{X-ray data originally measured in 0.12-2.48~keV.} We used radii derived from the stellar spectral types converted to stellar radius using the relations of \cite{Pecaut2013}. Only stars with published ages, X-ray luminosities, and a spectral type are shown in Fig.~\ref{fig:bootstrap_Lx}.
\end{itemize}

To constrain the spread in X-ray luminosity, we bootstrapped the data 1000 times; for each run we selected one point from each cluster in the data from \cite{Nunez2016}. We additionally selected one star from \cite{Booth2017} in each of three 0.5~dex age bins from log(Age) = 9 to 10.5. Once each older star was picked, an age and flux value was chosen for that star assuming a Gaussian error distribution.



\end{document}